\documentclass[aip,pof,reprint,onecolumn]{revtex4-1}
\usepackage{amsmath}
\usepackage{graphicx}
\usepackage{color}
\usepackage{hyperref}
\usepackage{psfrag}
\usepackage{stmaryrd}
\usepackage{amssymb}
\usepackage{wasysym}
\usepackage[latin1]{inputenc}

\def \degre {$^\mathrm{o}$}

\begin{document}

\title{Inertial waves and modes excited by the libration of a rotating cube}

\author{Jean Boisson}
\email{boisson@fast.u-psud.fr}
\affiliation{Laboratoire FAST, CNRS, Universit\'e Paris-Sud, UPMC, France}
\author{Cyril Lamriben}
\email{lamriben@fast.u-psud.fr}
\affiliation{Laboratoire FAST, CNRS, Universit\'e Paris-Sud, UPMC, France}
\author{Leo R. M. Maas}
\email{Leo.Maas@nioz.nl}
\affiliation{NIOZ Royal Netherlands Institute for Sea Research, Texel,
The Netherlands}
\author{Pierre-Philippe Cortet}
\email{ppcortet@fast.u-psud.fr}
\affiliation{Laboratoire FAST, CNRS, Universit\'e Paris-Sud, UPMC, France}
\author{Fr\'{e}d\'{e}ric Moisy}
\email{moisy@fast.u-psud.fr}
\affiliation{Laboratoire FAST, CNRS, Universit\'e Paris-Sud, UPMC, France}

\date{\today}

\begin{abstract}

We report experimental measurements of the flow in a cubic
container submitted to a longitudinal libration, i.e. a rotation
modulated in time. Velocity fields in a vertical and a horizontal
plane are measured in the librating frame using a corotating particle image
velocimetry system. When the libration frequency $\sigma_0$ is
smaller than twice the mean rotation rate $\Omega_0$, inertial
waves can propagate in the interior of the fluid.  At arbitrary
excitation frequencies $\sigma_0<2\Omega_0$, the oscillating flow
shows two contributions: (i) a basic flow induced by the libration
motion, and (ii) inertial wave beams propagating obliquely upward
and downward from the horizontal edges of the cube. In addition to
these two contributions, inertial modes may also be excited at
some specific resonant frequencies. We characterize in particular
the resonance of the mode of lowest order compatible with the
symmetries of the forcing,  noted [2,1,+]. By comparing the
measured flow fields to the expected inviscid inertial modes computed
numerically [L.R.M. Maas, Fluid Dyn. Res. \textbf{33}, 373 (2003)], we show
that only a subset of inertial modes, matching the symmetries of
the forcing, can be excited by the libration.

\end{abstract}

\maketitle

\section{Introduction}

Rotating fluids support the existence of a singular class of waves
called inertial
waves,\cite{Greenspan1968,Lighthill1978,Pedlosky1987} which are
anisotropic and propagate because of the restoring nature of the
Coriolis force. These waves exist only for excitation
frequencies $\sigma_0$ lower than twice the rotation rate
$\Omega_0$. In a confined volume of fluid, inertial waves may
become phase-coherent and experience a resonance due to multiple
reflections over the container walls, leading to the so-called
inertial modes. These inertial modes are relevant to geophysical
and astrophysical flows (in liquid cores of planets and stars),
but also to mechanical engineering flows (e.g., in liquid-filled
projectiles).

Inviscid inertial modes are the eigenmodes of a given container
geometry, and can be found in general when the walls are either
normal or parallel to the rotation
axis,\cite{Greenspan1968,McEwan1970} but also in some specific
cases such as spheres, spheroids\cite{Zhang2004} and, to some
extent, in spherical shells (namely for so-called R-modes that
lack  radial displacement). However, in general, when sloping
walls are present, the wave focusing and defocusing induced by the
peculiar reflection law of inertial waves\cite{Phillips1963}
precludes the existence of inviscid eigenmodes, and the
concentration of energy along particular beams leads to wave
attractors.\cite{Manders2003} The resonance frequencies of
inviscid inertial modes can be derived analytically only in some
specific geometries, such as cylinders, spheres and
spheroids.\cite{Batchelor1967,Greenspan1968,Zhang2004} In the case of a
parallelepipedic container, such as used  in the experiments presented here, the frequencies and the spatial
structure of the inviscid inertial modes have been determined
numerically.\cite{Maas2003}

In geophysical and astrophysical flows, several types of forcing
may be at the origin of the excitation of inertial modes. Inertial
modes have been excited experimentally in a
spherical cavity by a longitudinal libration\cite{Aldridge1969}
---i.e. a time modulation of the rotation rate---, and have
recently been described in numerical simulations in spheres and spherical
shells.\cite{Rieutord1991,Tilgner1999,Noir2009,Calkins2010}
Precession\cite{Busse1968,Kerswell1995,Noir2001} and periodic
deformation of the walls modeling gravitational
tides\cite{Suess1971,Morize2010} are other examples of forcing,
providing an efficient generation of inertial modes. These
inertial modes have been proposed to contribute, through
nonlinear self-interaction, to the generation of steady
zonal flows, which are visible for instance in the atmosphere
of gaseous planets like Jupiter.\cite{Malkus1968,Tilgner2007,Morize2010}

Although axisymmetric geometries have been primarily considered
to observe and characterize inertial modes, non-axisymmetric geometries such as parallelepipeds\cite{Maas2003}
are also of fundamental interest. 
In particular, the symmetries of the container, the symmetries of the forcing,
and the role of viscosity, are critical parameters
in determining which inertial modes can be excited.

Inertial modes may be present
in any laboratory experiment performed in a rotating container.
They have been for instance detected in
ensemble averages of turbulence generated by the translation
of a grid in a rotating
container.\cite{Dalziel1992,Bewley2007,Lamriben2011a} When present, these
modes may couple to the turbulence, and have a profound influence
on its statistical properties. In particular, turbulence in the
presence of inertial modes, which are non homogeneous by nature,
cannot be considered as freely decaying, raising the issue of the
relevance of the homogeneous framework to describe
rotating turbulence in laboratory experiments.

In this paper, we investigate the spatial structure of the
oscillating flow generated by the longitudinal libration of a
cube, and we characterize the efficiency of this configuration to
excite inertial modes. Measurements are performed both
in a vertical and in a horizontal plane (the rotation axis is
vertical) using a co-rotating two-dimensional
particle image velocimetry (PIV) system. For libration frequencies
lower than twice the rotation rate $\sigma_0<2\Omega_0$, we
observe, in addition to the basic oscillating flow induced by the
libration, the propagation of oblique inertial wave beams emitted
from the horizontal edges of the cube. These wave beams, similar
to the ones observed in cylindrical
geometry,\cite{McEwan1970,Duguet2006} originate from the
convergence of Ekman fluxes near the edges of the container.
In addition, for a specific libration frequency in the explored
range $\sigma_0 / 2\Omega_0 \in [0.60; 0.73]$, libration also excites
an inertial mode, of spatial structure in close agreement with the
inviscid computation of Ref.~\onlinecite{Maas2003}. Our results suggest that,
among all the eigen modes predicted numerically, only a small subset of
low order modes matching exactly the symmetries of the cube
libration can be actually excited.

\section{Experimental Setup}\label{sec:setup}

\subsection{Flow geometry and rotating platform}

\begin{figure}
    \centerline{\includegraphics[width=8cm]{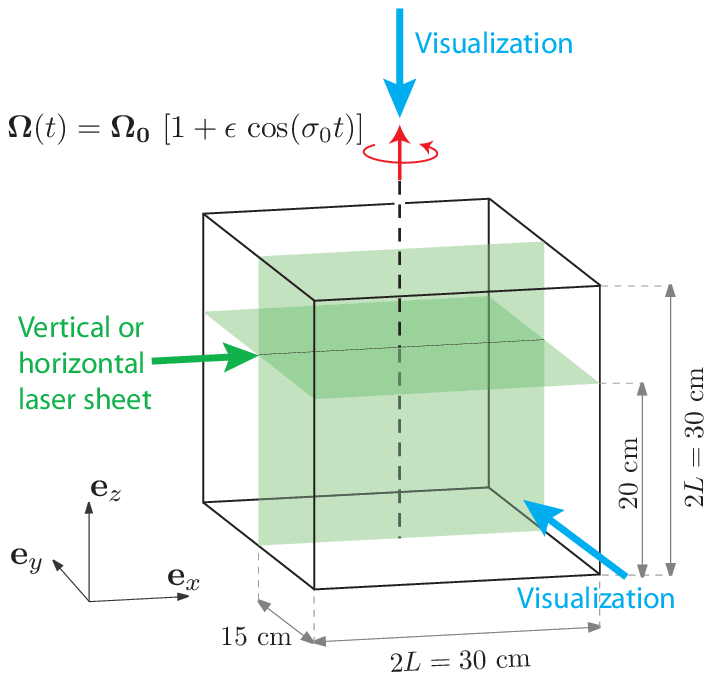}}
    \caption{(Color online) Schematic view of the (30~cm)$^3$ water
    tank mounted on the rotating platform. The bottom and top walls
    are located at $z=0$ and $z=2L$, and the origin of the coordinates
    $(x,y)$ is taken at the center of the square cross section.
    Two-dimensional PIV measurements are achieved in the vertical
    plane $y=0$ and in the horizontal plane $z=4L/3$  in the librated
    frame using a co-rotating laser sheet and a camera aiming normally
    at it. \label{fig:setup}}
\end{figure}

The experimental setup, sketched in Fig.~\ref{fig:setup}, consists
in a closed cubic glass tank, of inner size $2L=30$~cm, filled
with water and mounted on a precision rotating turntable of $2$~m
in diameter. We use a coordinate system centered on the horizontal
square, $-L \leq x,y \leq L$, with the horizontal walls at $z=0$
and $z=2L$, perpendicular to the rotation vector ${\bf
\Omega}=\Omega(t) {\bf e}_z$. The mean angular velocity of the
rotating platform is set to $\Omega_0=0.419$~rad~s$^{-1}$ (4 rpm).
The Ekman number of the system, which compares the viscous to the
Coriolis force, is $E =\nu/(2\Omega_0 L^2)=5.3 \times 10^{-5}$.
The rotation of the fluid is set long before the start of the
libration, at least $1$~h, in order for transient spin-up
recirculations to be damped. Once the solid-body rotation is
reached, the turntable is submitted to a longitudinal libration,
which consists in a modulation of the angular velocity $\Omega(t)$
around $\Omega_0$ at a frequency $\sigma_0$ with a peak-to-peak
amplitude $2 \epsilon\,\Omega_0$,
\begin{equation}
\Omega(t)=\Omega_0 \,\left[1 + \epsilon\, \cos(\sigma_0 t)\right].
\label{eq:libration}
\end{equation}
In the frame rotating at constant velocity $\Omega_0$,
the libration of the cube is described by the angle
\begin{equation}
\varphi(t) = \epsilon \frac{\Omega_0}{\sigma_0} \sin(\sigma_0 t).
\label{eq:angle}
\end{equation}

The normalized libration amplitude $\epsilon$, which is
the Rossby number of the problem, is varied between $2\%$ and
$16\%$. In order to excite inertial waves, the libration frequency
$\sigma_0$ can take in principle any value in the range $[0,
2\Omega_0]$. In this paper, we focus on the restricted range
$\sigma_0 / 2 \Omega_0 \in [0.60; 0.73]$. For these frequencies,
the peak-to-peak libration angle $\Delta \varphi = 2 \epsilon
\Omega_0 / \sigma_0$ lies in the range from $1^\mathrm{o}$ to
13$^\mathrm{o}$ for $\epsilon = 2\%$ to $16\%$. The relative
precision in the control of the instantaneous rotation rate is
better than $10^{-3}$. After the start of the libration, we wait
at least half an hour before the data acquisition (which
represents $8 \tau_E$, where $\tau_E = L (\nu \Omega_0)^{-1/2}$ is
the Ekman timescale) in order to reach a stationary regime.

\subsection{Particle image velocimetry (PIV) measurements}

Velocity fields are measured in the librated reference frame using
a two-dimensional PIV system mounted on the rotating platform
(Fig.~\ref{fig:setup}).\cite{Davis,pivmat} Measurements are
performed either in the vertical plane $y=0$ or in the horizontal
plane $z=4L/3=20$~cm. This last particular choice is motivated by
the fact that $z=4L/3$ does not correspond to a node for the
inertial modes considered in this paper (the centered horizontal
plane $z=L$ is a node for the inertial modes of even vertical
wavenumber $n$). The flow is seeded with $10$~$\mu$m tracer
particles, and illuminated by a corotating laser sheet generated
by a $140$~mJ Nd:YAG pulsed laser. For both horizontal and
vertical measurements, the entire $30\times30$~cm$^2$ flow
sections are imaged through transparent sides of the tank with a
high resolution $2048 \times 2048$~pixels camera aiming normally
at the laser sheet.

Each acquisition consists in at least 1000, and up to 3000, images
taken at a sampling rate between $12\,\sigma_0$ and $36\,\sigma_0$
which correspond to at least 80 libration periods. The sampling
rate is chosen according to the libration amplitude in order to
keep a typical particle displacement of the order of 5 pixels
between two successive images. PIV fields are computed over
successive images using $32 \times 32$ pixels interrogation
windows with $50\%$ overlap, leading to a spatial resolution of
$3.5$~mm. This resolution is not enough to resolve the thickness
of the Ekman boundary layers, $\delta_E = L \, E^{1/2} \simeq
1$~mm, but is appropriate for the flow structures associated to
inertial waves and modes in the bulk of the fluid.

\section{Basic libration flow}
\label{sec:baseflow}

\subsection{Inviscid solution}

We first characterize the basic flow response to the libration
forcing. Since in the present experiments the modulation period
$2\pi/\sigma_0 \simeq [20;25]$~s is much shorter than the Ekman time
$\tau_E \simeq 230$~s (which is the timescale required for the
angular velocity of the fluid to match that of the boundaries),
the core of the flow can be considered as essentially inertial,
and rotating at the constant rotation rate $\Omega_0$. In the
frame rotating at ${\Omega_0}$, the flow is therefore
approximately at rest, and surrounded by oscillating
recirculations induced by the periodic motion of the walls.

The inviscid response to the libration of an arbitrary container,
whose walls are either parallel or normal to the rotation axis,
can be derived by assuming that the flow is strictly
two-dimensional. In the absence of viscosity, the {\it absolute
vorticity} of the fluid in the frame of the laboratory,
$$
{\boldsymbol \omega}_a = {\boldsymbol \omega} + 2 \Omega(t) {\bf e}_z,
$$
must be conserved, and hence given by the mean vorticity $2
\Omega_0 {\bf e}_z$. Here ${\boldsymbol \omega}$ is the {\it
relative vorticity}, as measured in the libration frame. If
viscosity is present, this result remains approximately valid far
from the boundaries, with the additional assumption that $\sigma_0
/ \Omega_0 \gg E^{1/2}$ (i.e., for a rapid libration compared to
the Ekman timescale). With the total angular velocity of the
libration given by Eq.~(\ref{eq:libration}), the relative
vorticity ${\boldsymbol \omega}$ is therefore vertical,
homogeneous, and  given by
\begin{equation}
\omega_z(t)  ={\partial u_y \over \partial x}-{\partial u_x \over \partial y}= - 2 \epsilon \Omega_0 \cos (\sigma_0 t).
\label{eq:wzt}
\end{equation}
In an axisymmetric container, the resulting flow is an oscillating
solid body rotation of angular velocity $-\epsilon \Omega_0 \cos
(\sigma_0 t)$: this simply describes a fluid at rest in the frame
rotating at constant rate $\Omega_0$. The case of a
non-axisymmetric container is more complex, and can be solved in
terms of a streamfunction $\psi(x,y,t)=\Psi(x,y) \cos(\sigma_0
t)$, such that $(u_x,u_y)=(-\partial_y \psi, \partial_x \psi)$.
Equation (\ref{eq:wzt}) therefore takes the form of a Poisson
equation for the spatial part of the streamfunction,
$$
\Delta \Psi = -2 \epsilon \Omega_0.
$$
The solution of this equation for a square domain, subject to the
condition that the streamfunction vanishes at the boundary, is
given in the Appendix, in Eq.~(\ref{eq:psisol}). The streamlines
are circular near the center of the container, as in a solid-body
rotation, but they become gradually square near the boundaries
(see the velocity field in Fig.~\ref{fig:base}a).

\subsection{Experimental measurements of the libration flow}

\begin{figure}
    \centerline{\includegraphics[width=11cm]{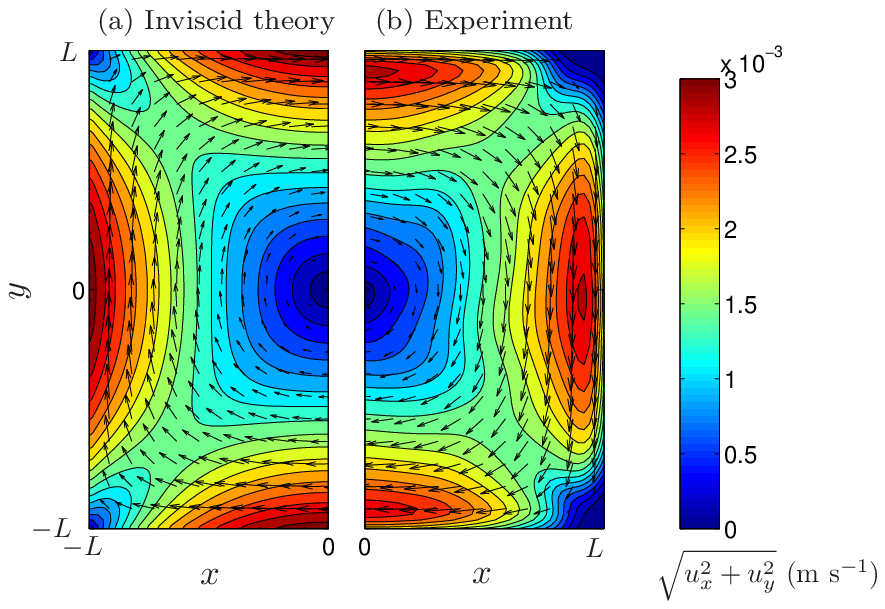}}
    \caption{(Color online) Velocity fields excited by libration in a
    container of square cross-section. (a) Inviscid solution, computed
    from Eq.~(\ref{eq:psisol}). (b) PIV measurements at the phase
    $\sigma_0 t = 0$ (maximum libration velocity). The libration
    frequency is $\sigma_0 / 2\Omega_0 = 0.648$ (corresponding to the
    mode $[2,2,+]$, which is not resonant in this experiment), and its
    amplitude is $\epsilon = 0.04$. \label{fig:base}}
\end{figure}

Figure~\ref{fig:base} compares the measured libration flow
obtained by PIV and the inviscid solution (\ref{eq:psisol}). The
velocity field shown here is taken at the phase $\sigma_0 t = 0$
corresponding to the maximum amplitude of the libration, when the
flow in the librated frame is anticyclonic (i.e., $\omega_z = - 2
\epsilon \Omega_0$). The libration frequency $\sigma_0$ is chosen
here far from any resonant frequency of inertial mode. The
agreement between the measured field and the inviscid solution is
excellent in most of the flow section, except near the boundaries.
It must be noted that the vanishing of the velocity near the
boundaries may be affected by the PIV resolution (about 3~mm),
which is of the same order as the expected boundary layer thickness.

An interesting feature of Fig.~\ref{fig:base}(b) is the wavy shape
of the experimental iso-velocity lines: this is a first indication that, in
addition to the basic libration flow, the flow also contains a
wave component. This additional component is further described in
the next Section.

\begin{figure}
    \centerline{\includegraphics[width=8cm]{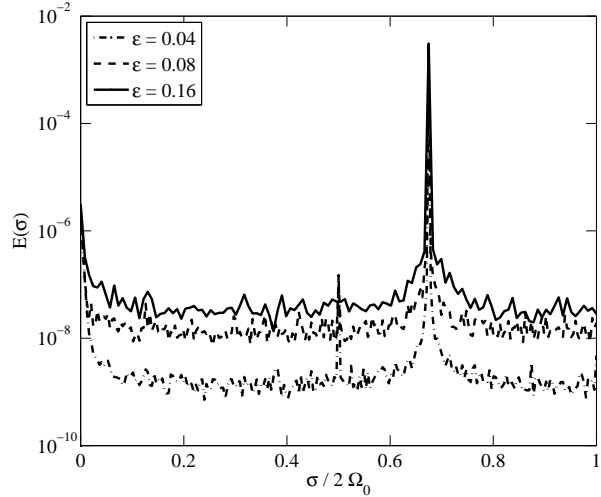}}
    \caption{Temporal energy spectra $E(\sigma)$ for a forcing
    frequency $\sigma_0 / 2 \Omega_0 = 0.675$ (corresponding to the
    mode $[2,1,+]$), for three libration amplitudes $\epsilon = 0.04,
    0.08$ and 0.16. \label{fig:spectrum}}
\end{figure}

For all the libration amplitudes investigated here ($\epsilon =
2\%$ to $16\%$), the fluid response to the forcing is found to
remain essentially linear. This is demonstrated in
Fig.~\ref{fig:spectrum}, where the temporal energy spectra
$E(\sigma)$ are shown for different forcing amplitude $\epsilon$.
Here, the spectrum is defined as $E(\sigma) = \langle | \hat{\bf
u}_\sigma |^2 \rangle_{xy}$, where $\hat{\bf u}_\sigma(x,y)$ is
the temporal Fourier transform computed at each location, and
$\langle \rangle_{xy}$ is the spatial average over the horizontal
plane.  The Fourier transform is computed over at least 80 periods
of libration.

All spectra show a narrow peak at the forcing frequency
$\sigma_0$, well above the white noise level of the PIV
measurements. We note that, since we are only interested here in
forcing frequencies $\sigma_0 > \Omega_0$, higher harmonics
$n\sigma_0$, if present, are beyond the upper limit $2 \Omega_0$,
and are therefore not governed by the dynamics of inertial waves.

We can note that, for all forcing amplitudes $\epsilon$, the peak
at $\sigma_0$ contains at least 97\% of the total flow energy. The
remaining energy is associated to secondary peaks at $\sigma = 0$,
$\sigma = \Omega_0$ and higher harmonics ($\sigma > 2\Omega_0$).
The secondary peak at $\sigma = \Omega_0$, mostly visible at low
$\epsilon$, corresponds to a residual fluid motion synchronized
with the platform rotation rate. The peak at $\sigma = 0$ can be
associated to the generation of a weak mean flow. This mean
flow is probably a nonlinear response to the libration forcing,
and is not investigated in the present paper.

In order to improve the signal-to-noise ratio, we perform in the
following a band-pass filtering of the velocity fields at the
forcing frequency $\sigma_0$. This procedure consists in filtering
the Fourier transform $\hat {\bf u}_\sigma$ at all $\sigma$ except
in a narrow region centered on $\sigma_0$ of width $\delta \sigma
/ 2 \Omega_0 = 0.024$ (which contains the energy of the peak to
within 1\%), and computing the inverse Fourier transform of the
resulting filtered field.

\section{Inviscid inertial modes and symmetries of the libration forcing}

\subsection{Inviscid inertial modes}

The eigen frequencies and the spatial structures of the inertial
modes in a parallelepiped whose sides are parallel or
perpendicular to the rotation axis have been determined
numerically for an inviscid fluid in Ref.~\onlinecite{Maas2003}.
These predictions have been achieved from the numerical resolution
of the eigenproblem defined from the inviscid linearized equations
in a rotating frame. Any frequency $\sigma_0$ in the range $[0,
2\Omega_0]$ corresponds to an inertial mode. Accordingly, an
infinity of inviscid modes may be excited if the system is forced
at a given frequency with a finite bandwidth.

Following the notation introduced in
Ref.~\onlinecite{Lamriben2011a}, we label the inertial modes as
[$n,m,s$]. The first index $n$ is the normalized vertical
wavenumber, such that the velocity field ${\bf u}$ shows $n$
recirculation cells in the vertical direction. With the container
walls taken at $z = 0$ and $z=H$, the velocity field has the form
\begin{eqnarray}
{\bf u}_\perp(x,y,z,t)&=&\cos \left(\pi n \frac{z}{H}\right)\,\breve{\bf u}_\perp(x,y,t),\\
u_z(x,y,z,t)&=&\sin \left(\pi n \frac{z}{H}\right)\,\breve{u}_z(x,y,t),
\end{eqnarray}
with ${\bf u}_\perp = u_x {\bf e}_x + u_y {\bf e}_y$
(we restrict to
$H=2L$ in this paper). Inertial modes are therefore stationary in
the vertical direction, but their horizontal structure $\breve{\bf
u}(x,y,t)$ may be either stationary (the so-called ``sloshing''
modes) or propagating. The second index $m$ enumerates the
eigen frequencies of modes of sign $s$ from the largest one ($m=1$)
down. Larger values of $m$ essentially correspond to finer
structures in the horizontal plane.  Since each mode $m$ is
expressed in terms of an infinite amount of horizontal Fourier
modes, the index $m$  is not directly related to a number of nodes
as for the vertical index $n$. Finally, the sign $s$ refers to the
invariance of the mode with respect to rotation of angle $\pi$
about the $z$ axis. More precisely, a mode ${\bf u}(x,y,z,t)$ has
symmetry $s$ if
\begin{equation}
\left( \begin{array}{c} u_x \\ u_y \\ u_z \end{array} \right)
(-x,-y,z,t) = s \left( \begin{array}{c} -u_x \\ -u_y \\ u_z
\end{array} \right) (x,y,z,t). \label{eq:defsym}
\end{equation}

The resonance frequencies of inertial modes $[n,m,s]$ are
increasing functions of $n$ and, at fixed $n$ and $s$, decreasing
functions of $m$. This behavior essentially originates from the
dispersion relation for plane inertial waves,
\begin{equation}
\sigma = 2\Omega_0 \cos \theta,
\label{eq:rd}
\end{equation}
where $\theta$ is the angle between the wavevector ${\bf k}$ and
the rotation axis, $\cos \theta = k_z / |{\bf k}|$. Low
frequencies $\sigma$ are therefore associated to nearly horizontal
${\bf k}$, i.e. to small $k_z$ and hence to small $n$, or/and to
large $k_{x,y}$ and hence to large $m$.

\subsection{Symmetry properties and boundary conditions for a viscous fluid}

In the case of a viscous fluid, the no-slip boundary condition at
the walls is expected to affect the spatial structure of the
inertial modes. The spectrum and spatial structure of viscous
inertial modes in a rotating parallelepiped have not been computed yet.
It is therefore of first interest to check to what extent the
inviscid modes found numerically could be recovered in a viscous fluid at finite
Ekman number. Since viscosity damps preferentially the high order
modes (i.e. the modes with large indices $n$ and/or $m$), we
expect the libration to force more efficiently modes of low order.
This is straightforward for the vertical wavenumber $n$, which is
naturally associated to a viscous damping proportional to $\nu
n^2$. No equivalent simple law exists for the horizontal index
$m$, but we can similarly expect that lower $m$, associated to
larger scales in the horizontal plane, will be favored in the
presence of viscosity.

From the symmetries of the boundary conditions, it is possible to
anticipate which inertial modes are compatible with the libration
forcing. In the frame rotating at constant velocity $\Omega_0$,
the angular oscillation of the top and bottom walls is described
by the velocity ${\bf u}(x,y,z=0) = {\bf u}(x,y,z=2L) = \epsilon
\Omega_0 \cos(\sigma_0 t)\,(-y {\bf e}_x + x {\bf e_y})$ which,
according to Eq.~(\ref{eq:defsym}), has symmetry $s=+$. Moreover,
the no-slip boundary conditions impose equal horizontal velocity
at the top and bottom walls, which is satisfied only for even
vertical wavenumbers $n$ (the basic libration flow described in
the previous section is vertically invariant, and is hence
associated to $n=0$). Finally, there is the possibility that
particular symmetries associated to the second index $m$ may even
further reduce the set of modes compatible with the symmetries of
the libration forcing. According to these symmetry properties, the
longitudinal libration is expected to excite only a subset of the
modes $[n,m,s]$ among those having even $n$ and $s=+$.

It must be noted that those symmetry arguments state which modes
are forbidden by this forcing, but they do not state which modes,
among the allowed ones, are effectively excited by the libration.
In other words, we would like to address the question of how the
fluid motion induced by the oscillating walls, and in particular
close to the edges of the cube, will be transmitted to the
inertial modes, and to characterize the efficiency of this
transmission.

\begin{table}
\begin{tabular}{p{2.5cm}p{1.6cm}}
\hline \hline
 mode &  frequency \\
$[n,m,s]$ & \hspace{0.5cm}$\sigma/2\Omega_0$ \\
\hline
$[2,1,+]*$ & \hspace{0.5cm}0.6742\\
$[2,2,+]*$ & \hspace{0.5cm}0.6484\\
\hline
$[3,2,-]$ & \hspace{0.5cm}0.7271\\
$[3,3,-]$ & \hspace{0.5cm}0.6857\\
$[3,4,-]$ & \hspace{0.5cm}0.6848\\
$[3,3,+]$ & \hspace{0.5cm}0.6485\\
$[3,4,+]$ & \hspace{0.5cm}0.6258\\
\hline
$[4,5,+]*$ & \hspace{0.5cm}0.6960\\
$[4,6,+]*$ & \hspace{0.5cm}0.6945\\
$[4,5,-]$ & \hspace{0.5cm}0.6889\\
$[4,6,-]$ & \hspace{0.5cm}0.6643\\

\hline \hline
\end{tabular}
\caption{List of all the inviscid inertial modes $[n,m,s]$,
truncated at $n \leq 4$ and $m \leq 6$, in the range of
frequencies $\sigma_0/2\Omega_0 \in [0.60; 0.73]$. $n$ is the
normalized vertical wavenumber, $m$ characterizes the horizontal
structure, and $s=\pm$ refers to the symmetry or antisymmetry of
the mode with respect to rotation of angle $\pi$ about the
rotation axis (see Ref.~\onlinecite{Maas2003} for details). The *
symbol marks the 4 modes having symmetries compatible with the
libration forcing: even $n$ and $s=+$.
 \label{tab:1}}
\end{table}

In the following, we investigate the flow response for libration
frequencies in the vicinity of two inertial modes of low order
allowed by the symmetries of the forcing: [2,1,+] and [2,2,+].
More specifically, we systematically characterize the flow in the
range  $\sigma_0 / 2\Omega_0 \in [0.60; 0.73]$ surrounding these
two modes. The list of all modes in this range, truncated
at $n \leq 4$ and $m \leq 6$, is given in Tab.~\ref{tab:1}. Among
these modes, only the four ones [2,1,+], [2,2,+], [4,5,+] and
[4,6,+] (marked by a * symbol) have the correct symmetries (even
$n$ and $s=+$) and could be observed in principle. However, we
will show in the following that viscous effects and additional
symmetry properties actually select only the single mode [2,1,+]
in this list.

\section{Edge beams and inertial modes}

\subsection{Flow in the vertical plane: non-resonant case}

\begin{figure}
    \centerline{\includegraphics[width=13cm]{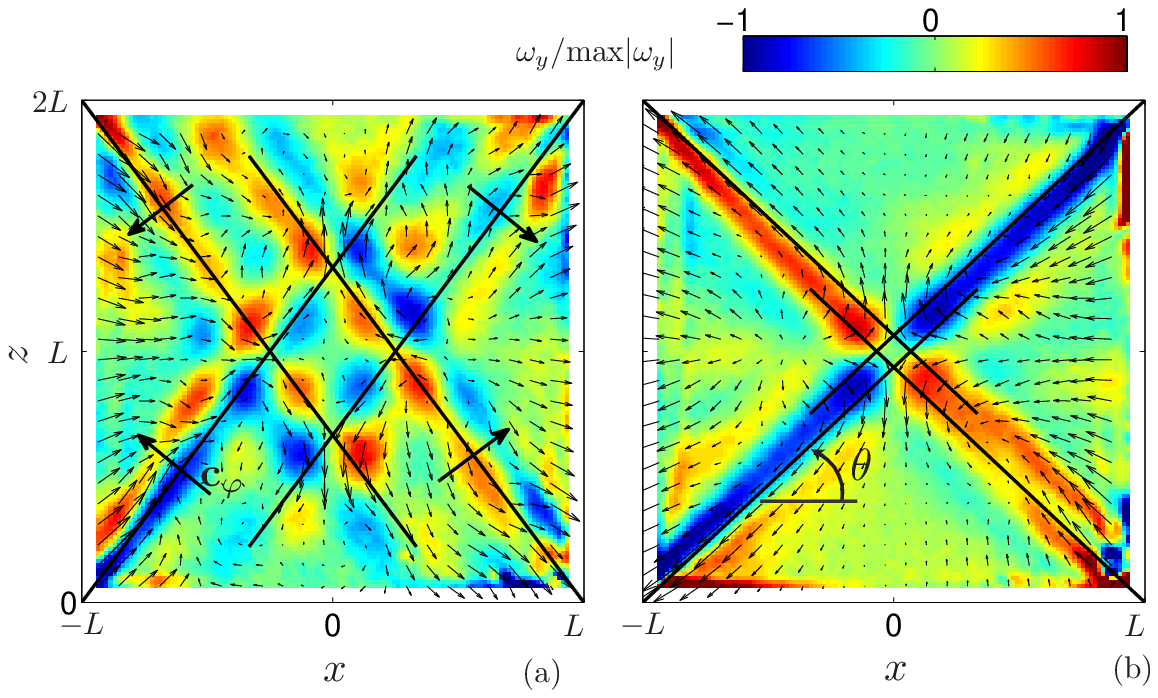}}
    \caption{(Color online) Spatial structure, in the  vertical plane
    $y=0$, of the oscillating flow excited with $\epsilon=0.02$ and at
    frequencies $\sigma_0/2\Omega_0=0.60$ (a) and $0.73$ (b), in the
    absence of inertial modes. The velocity fields are extracted by
    temporal band-pass filtering around the excitation frequency. The
    arrow fields illustrate the $(x,z)$ components of the velocity
    field at a given arbitrary phase of the oscillation, and the gray
    shade/color maps the corresponding normalized vorticity
    $\omega_y/{\rm max}|\omega_y|$. Resolution of the velocity fields
    has been reduced by a factor 5 for better visibility. In (a), bold
    arrows show the direction of the  phase velocity ${\bf c}_\varphi$
    for each wave beam. In (a) and (b), the solid black lines show the
    expected direction predicted for each wave beam from the
    dispersion relation, $\cos\theta=\sigma_0/2\Omega_0$.
    \label{fig:croix}}
\end{figure}

We first describe the flow in the vertical plane $y=0$ in the
absence of inertial modes. The velocity ${\bf u}_{x,z}$ and the
vorticity $\omega_y(x,z)$  shown in Fig.~\ref{fig:croix} for
$\epsilon=0.02$ are for forcing frequencies
$\sigma_0/2\Omega_0=0.60$ (a) and $0.73$ (b), chosen far from any
low order inertial mode. Fields are not displayed in lateral
strips of 12~mm width along the walls due to poor quality of the
data induced by light reflections. Since the basic libration flow
is normal to the plane $y=0$,  the vertical flows here show the
$x$ and $z$ deviations from the libration flow, and contain
therefore only the contributions from the wave component of the
flow.

We observe that the libration excites a cross shaped pattern made
of four oblique shear layers originating from the bottom and top
edges of the cube. These shear layers actually correspond to
propagative inertial waves in two-dimensional plane wave packets,
which contain approximately one
wavelength in their transverse direction. There are actually
eight two-dimensional wave packets emitted from the four upper and the four
lower edges of the container; only the four ones emitted from
the edges defined by $x = \pm L$, $z=0,2L$ can be seen in the vertical cut $y=0$.
The direction of the beams is found
in good agreement with the dispersion relation (\ref{eq:rd}), as
shown by the black lines in Fig.~\ref{fig:croix} making an angle
$\theta= \cos^{-1}(\sigma_0/2\Omega_0)$ with the horizontal. In
these wave beams,  fluid particles describe anticyclonic circular
motion at frequency $\sigma_0$ in the planes inclined at angles
$\pm \theta$ with the horizontal (the sign depends on the
considered beam). The shearing motion traced by the vorticity
$\omega_y$ is related to the variation of the wave phase in the
direction normal to $\pm \theta$. We also observe that the phase
of the wave travels normal to the beam, with a phase velocity
${\bf c}_\varphi$ directed towards the vertical wall from which
originates the beam. In Fig.~\ref{fig:croix}(a)
[$\theta=\cos^{-1}(0.60)\simeq 53$\degre], the four wave beams
show interferences at their intersections, which make the wave
pattern rather complex. On the contrary, in
Fig.~\ref{fig:croix}(b) [$\theta=\cos^{-1}(0.73)\simeq 43$\degre],
the opposite wave beams are almost aligned for this particular
frequency. As a consequence, the wave beams show constructive
interferences, leading to almost standing waves along the
diagonals, but propagative outwards on their sides.

These edge beams are similar to the ones observed in cylindrical
containers, e.g. in the early experiment of
McEwan\cite{McEwan1970} (forced by a tilted top lid) and in the
numerical simulations of Duguet {\it et al.}\cite{Duguet2006}
(forced by an oscillating axial compression). Similar beams are
also found in the recent simulations of Sauret {\it et
al.}\cite{Sauret2012} (forced by a longitudinal libration),
although in this case the beam angle is apparently not directly
related to the libration frequency, and probably results
from the turbulence in the boundary layers over the sidewalls.
In all these references, the
container is cylindrical, resulting in conical edge beams, emitted
from the two circular edges and focusing towards the rotation
axis. In our non-axisymmetric geometry, since the disturbance
sources are linear segments, there are eight two-dimensional edge
beams emitted from the 4 upper and the 4 lower edges of the
container.

These edge beams originate from the oscillating outward (resp.
inward) Ekman layers on the upper and lower walls associated with
the prograde (resp. retrograde) part of the libration motion. Far
from the lateral walls, where the streamlines of the libration
flow are approximately axisymmetric, this flow is approximately
radial, of order $\epsilon \Omega_0 \sqrt{x^2+y^2}$. In the
classical axisymmetric spin-up problem, the horizontal flow in the
unsteady (but slowly varying) Ekman layer is vertically deflected
into Stewartson layers along the lateral
walls.\cite{Greenspan1968} However, in the rapidly oscillating
situation examined here, the matching condition between the Ekman
and Stewartson layers cannot be satisfied, and the flow in the
Ekman layer detaches when it reaches the lateral walls, generating
an oscillating shear layer in the bulk,\cite{Kerswell1995} of
characteristic velocity $\epsilon \Omega_0 L$. As a consequence, although
the velocity scale of the edge beams is inertial, viscosity is
responsible for this flow feature.

The width of these edge beams can be compared to predictions from
a viscous theory. Because of viscous damping, it is known that a
wave beam spreads as $\ell^{1/3}$, where $\ell$ is the distance
from the source.\cite{Peat1978,Tilgner2000} More precisely, the
thickening of the full-width half maximum $\delta(\ell)$ of a
self-similar plane beam emitted by a linear disturbance of small
size can be computed,\cite{Cortet2010}
\begin{eqnarray}
\delta(\ell) \simeq 6.84 L \left( \frac{E}{\sin \theta}
\right)^{1/3} \left( \frac{\ell}{L} \right)^{1/3},
\end{eqnarray}
with $E = \nu / (2 \Omega_0 L^2)$. According to this model, a wave
beam emitted from an edge reaches the central area of the
container after a distance $\ell^* = L / \sin \theta$ (for $\theta
> 45$\degre), where its width is $\delta(\ell^*) \simeq 6.84 (\sin
\theta)^{-2/3} \, L \, E^{1/3} \simeq 4.2-5.0$~cm for the range of
angles considered here. This prediction provides correct agreement
with the observations in Fig.~\ref{fig:croix}, showing beams of
thickness of order of $3-4$~cm near the center. Note that the case
$\theta \simeq 45$\degre~ is specific: the overlap of the upper
and lower wave beams leads to a different
scaling,\cite{Duguet2006} $\delta \simeq E^{1/4}$.

\subsection{Flow in the vertical plane: resonant case}

\begin{figure}
    \centerline{\includegraphics[width=14cm]{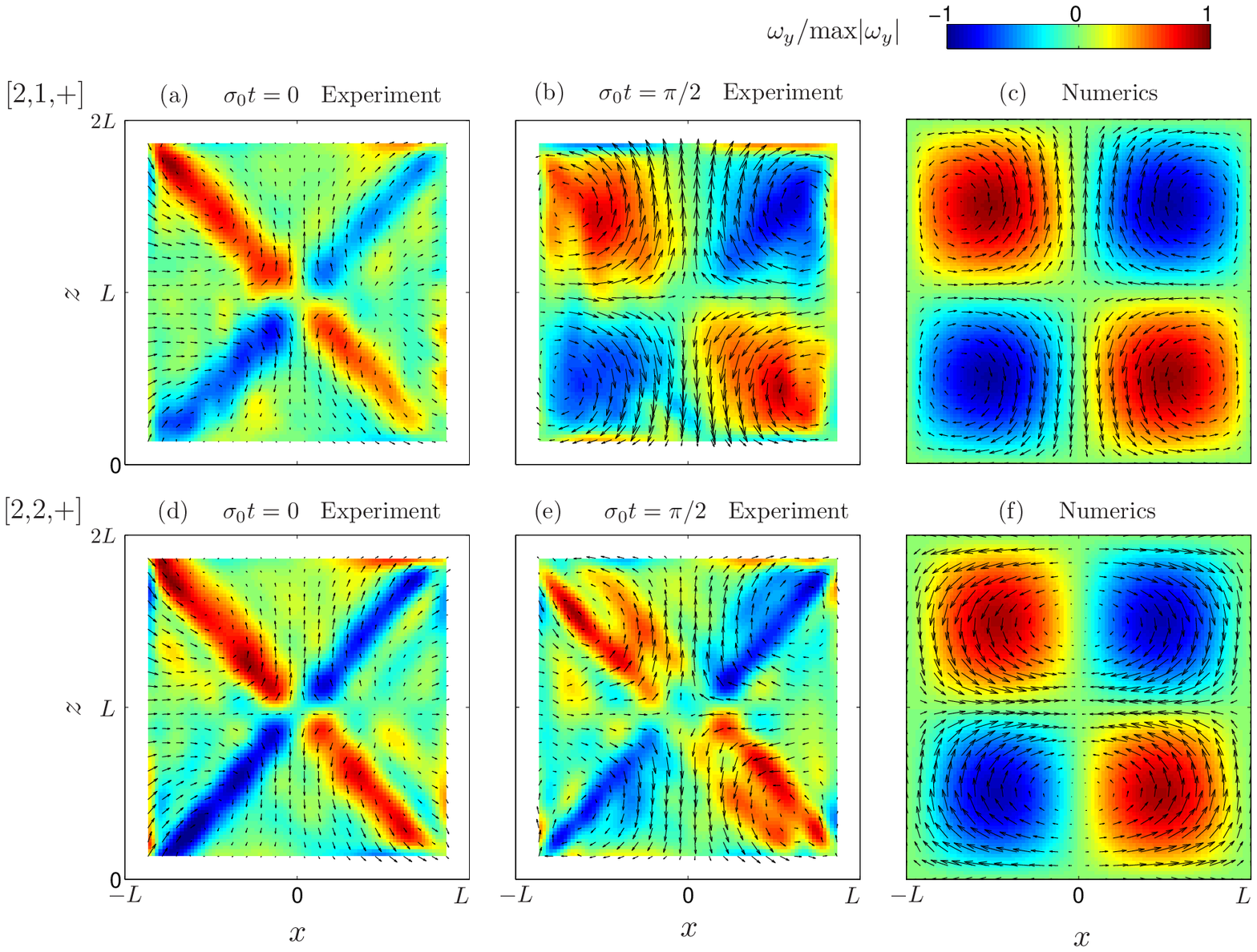}}
    \caption{(Color online) Spatial structure, in the vertical plane
    $y=0$, of the oscillating flow excited with $\epsilon=0.04$ and at
    frequency $\sigma_0/2\Omega_0=0.674$ (a-c) and
    $\sigma_0/2\Omega_0=0.648$ (d-f). (a,b,d,e) correspond to the
    experimental fields extracted by temporal band-pass filtering
    around the excitation frequency. (c,f) show the corresponding
    numerical inviscid modes [2,1,+] and [2,2,+] expected at the same
    frequency. The libration phase is $\sigma_0 t=0$ for (a,d), and
    $\sigma_0 t=\pi/2$ for (b,e). The gray shade/color maps the
    normalized vorticity field $\omega_y/{\rm max}|\omega_y|$.
    Resolution of the velocity fields has been reduced by a factor 5
    for better visibility (see also Ref.~\onlinecite{Suppl}).\label{fig:oscflowstruct}}
\end{figure}

We still examine here the structure of the flow in the vertical
plane, but now at frequencies where inertial modes are expected.
We select the two frequencies $\sigma_0/2\Omega_0=0.674$ and
$0.648$, corresponding to the inviscid predictions for the modes
[2,1,+] and [2,2,+] respectively (see Tab.~\ref{tab:1}). The
velocity and vorticity fields are shown in
Fig.~\ref{fig:oscflowstruct}, at the phases $\sigma_0 t=0$ (a,d)
and $\pi/2$ (b,e) of the libration. In the frame rotating at
constant velocity $\Omega_0$, the phase $\sigma_0 t=0$ corresponds
to $\varphi = 0$ and $\Omega(t) = \epsilon \Omega_0$ (i.e.,
maximum rotation rate), whereas $\sigma_0 t=\pi/2$ corresponds to
$\varphi$ maximum and $\Omega(t) = 0$ [see
Eqs.~(\ref{eq:libration})-(\ref{eq:angle})].

Looking first at the velocity fields (b) and (e), both frequencies
show two oscillating cells in the vertical direction, as expected
for a mode $n=2$ (see also the movies available at Ref.~\onlinecite{Suppl}).
These oscillating cells are mainly visible at
the phase $\sigma_0 t = \pi/2$, when the angle $\varphi$ is
maximum and the libration motion vanishes. At this phase, the
velocity fields clearly satisfy the $s=+$ symmetry, with
$u_{x}(-x,z) = -u_{x}(x,z)$ and $u_{z}(-x,z) = u_{z}(x,z)$. On the
other hand, at the phase $\sigma_0 t = 0$ corresponding to the
maximum libration velocity (a,d), the $x$ and $z$ components of
the velocity associated to this mode $n=2$ vanish and one recovers
the propagative edge beams already evidenced in
Fig.~\ref{fig:croix}.

Looking now at the vorticity field reveals finer structures of the
flow. At phase $\sigma_0 t = 0$ (a,d), the typical cross shaped
pattern composed of four edge beams is found for both forcing
frequencies. On the other hand, at phase $\pi/2$ (b,e), the shape
of the vorticity field is found to depend now on the forcing
frequency. Whereas it keeps approximately its cross-shape
structure for $\sigma_0/2\Omega_0 = 0.648$ (e), it shows four
nearly circular extrema of alternate signs for $\sigma_0/2\Omega_0
= 0.674$ (b). This indicates that, at $\sigma_0 t = \pi/2$, the
four recirculation cells dominate the vorticity for
$\sigma_0/2\Omega_0 = 0.674$ (b), whereas their vorticity is partially
hidden by the strong edge beams for $\sigma_0/2\Omega_0 = 0.648$
(e).

In order to understand the different behaviors between the two
frequencies, we compare now  the experimental fields with the
numerical predictions of the inviscid modes [2,1,+] and [2,2,+],
shown in Figs.~\ref{fig:oscflowstruct}(c) and (f), respectively
(movies of the numerical modes are also provided at Ref.~\onlinecite{Suppl}).
Note that the phase of the numerical fields is arbitrary since they
are not related to any specific forcing:
the numerical fields in Fig.~\ref{fig:oscflowstruct} have been simply
chosen here at phases which provide a
good visual correspondence with the experimental fields.
Although the vorticity patterns of modes [2,1,+] and [2,2,+]
look similar, with four extrema of alternate signs, their velocity
fields are very different: the vertical velocity is maximum on the
vertical axis for [2,1,+], but it is zero for [2,2,+]. The smooth
pattern of $\omega_y$ found experimentally in
Fig.~\ref{fig:oscflowstruct}(b) matches actually well the
prediction for [2,1,+]. On the other hand, the comparison fails
for Fig.~\ref{fig:oscflowstruct}(e), both for the velocity and
vorticity fields, suggesting that the mode [2,2,+] cannot be
excited by the longitudinal libration.

We can interpret these observations as follows. For
$\sigma_0/2\Omega_0=0.674$, the flow is a superimposition of the
pattern of propagative edge beams and a resonant stationary
[2,1,+] mode. Whenever the libration angle is zero (i.e. at phases
$\sigma_0 t=0$ and $\pi$), the instantaneous amplitude of the
[2,1,+] mode vanishes, and only the edge beams remain visible.
Turning now to the case $\sigma_0/2\Omega_0=0.648$, the mode
[2,2,+] is apparently not excited, and the edge beam pattern
remains visible all the time. Interestingly, it seems that
a flow reminiscent of the mode [2,1,+] is excited instead (see
Fig.~\ref{fig:oscflowstruct}(e)),
although with a weak amplitude,
probably because of the closeness of the eigen frequencies (less
than 4\%) of the two modes.  It will indeed be confirmed in
Sec.~\ref{sec:resonance} that the resonance of the [2,1,+] mode
spreads over a significant frequency range because of viscosity.

\subsection{Flow in the horizontal plane}

\begin{figure}
    \centerline{\includegraphics[width=14cm]{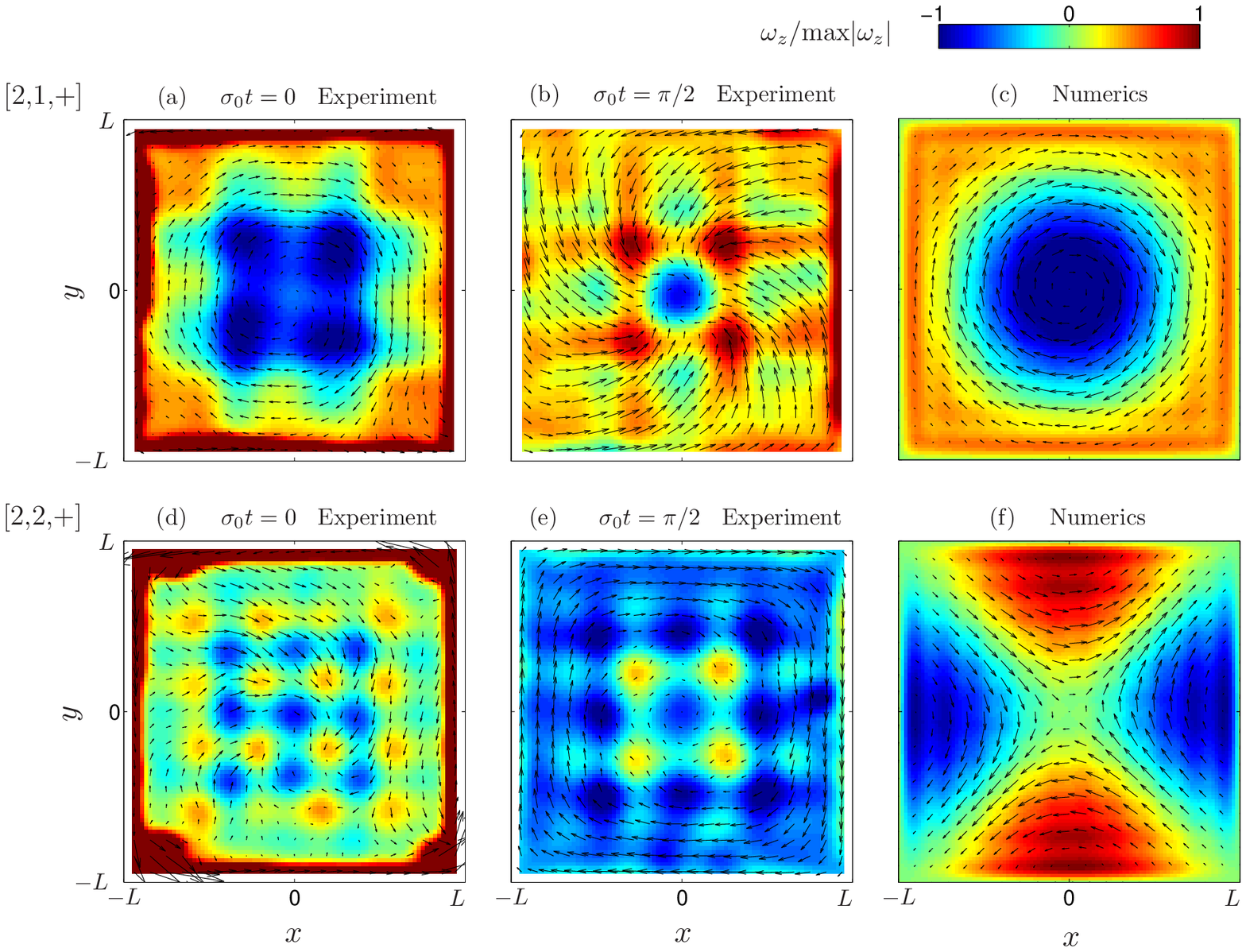}}
    \caption{(Color online) Spatial structure, in the horizontal plane
    $z=4L/3$, of the wave component of the flow ${\bf u'}$
    excited with $\epsilon=0.04$ and
    at frequency $\sigma_0/2\Omega_0=0.674$ (a-c) and
    $\sigma_0/2\Omega_0=0.648$ (d-f). Same layout as in
    Fig.~\ref{fig:oscflowstruct}.} \label{fig:rot22p_max}
\end{figure}

We finally turn to visualizations in the horizontal plane
$z=4L/3$, still for the two frequencies $\sigma_0/2\Omega_0=0.674$
and $0.648$. In order to focus on the wave component of the flow,
we remove in the following the libration component,
\begin{equation}
{\bf u'} (x,y,t) = {\bf u} (x,y,t) - {\bf u}_{lib}(x,y) \cos(\sigma_0 t),
\label{eq:umulib}
\end{equation}
where ${\bf u} (x,y,t)$ is the total velocity measured by PIV, and
${\bf u}_{lib}(x,y) = \nabla \times (\Psi {\bf e}_z)$ is the
inviscid libration flow computed from Eq.~(\ref{eq:psisol}). In
Fig.~\ref{fig:rot22p_max}, we show the resulting horizontal
velocity $u'_{x,y}$ and the corresponding vertical vorticity
$\omega'_z(x,y)$ fields for both frequencies, for the same phases
$\sigma_0 t = 0$ and $\pi/2$ as previously. Here again,
measurements are not displayed in lateral strips of width 6~mm
along the walls because of the poor quality of the PIV data.

For both  frequencies, the horizontal flow shows a complex spatial
structure which verifies the relation $u'_{x,y}(-x,-y) =
-u'_{x,y}(x,y)$, confirming the $s=+$ symmetry of the flow. Both
frequencies show a clear square pattern of vorticity extrema,
which is the trace of the interferences between the 8 propagative
inertial wave beams emitted from the 8 horizontal edges of the
container. Note that since the libration flow has constant
vorticity far from the boundaries, one has $\omega_z' = \omega_z +
2 \epsilon \Omega_0 \cos(\sigma_0 t)$, so the vorticity patterns
computed from the total flow and from its wave component are
identical. The amplitude of the wave component in the bulk of the
flow is typically a factor 10 below the libration velocity scale
$\epsilon \Omega_0 L$, indicating a weak efficiency of the
excitation of inertial waves by libration.

In Figs.~\ref{fig:rot22p_max}(c) and (f), we show for comparison
the numerical predictions of the inviscid modes
[2,1,+] and [2,2,+] in the horizontal plane.
The mode [2,1,+] (c) consists in a
nearly axisymmetric flow, in which each velocity vector
describes an elliptic oscillation in the anticyclonic direction.
During one period of the mode, the flow goes through
the following sequence: cyclonic, centrifugal, anticyclonic,
centripetal (see Ref.~\onlinecite{Suppl}). The mode [2,2,+] (f)
has a very different structure: it has four vortices located along
the sidewalls, with a hyperbolic point in the middle. This pattern
is essentially rotating as a whole around the rotation axis, with
only weak deformations near the walls due to the non-axisymmetry
of the container.

The comparison between the numerical inviscid modes and the
experimental measurements is revealing. The structure of the mode
[2,1,+] can be easily recognized in the experimental data,
superimposed to the square pattern of the interfering edge waves.
For instance, the numerical mode chosen at the phase shown in
Fig.~\ref{fig:rot22p_max}(c) matches well the experimental flow at
the libration phase $\sigma_0 t = 0$ [Fig.~\ref{fig:rot22p_max}(a)].
At this phase, the flow is normal to the plane $y=0$, in
agreement with the vanishing of the mode in the vertical plane
observed at the same phase in Fig.~\ref{fig:oscflowstruct}(a). On the
other hand, the structure of the inviscid mode [2,2,+] cannot be
recognized in the experimental data at any phase,\cite{Suppl} confirming that
this mode cannot be excited by libration. For both frequencies,
the vorticity field is essentially dominated by the interfering
edge waves, and is always different from the vorticity of the
predicted inviscid modes [Fig.~\ref{fig:rot22p_max}(c) and (f)].

The absence of mode $[2,2,+]$ in the experimental fields may be
understood from symmetry arguments. In addition to the $s=+$
symmetry, which corresponds to an invariance under a rotation of
$\pi$ about the $z$ axis, the [2,1,+] mode turns out to be also
invariant under a rotation of $\pi/2$ about the $z$ axis
[Fig.~\ref{fig:rot22p_max}(c)], which exactly matches the symmetry
of the libration forcing for a cube. On the contrary, the mode
[2,2,+] has only the $\pi$ symmetry, and it is antisymmetric with
respect to the rotation of $\pi/2$ about $z$. Taking into account
this additional symmetry explains why only [2,1,+] is excited by
the libration and not [2,2,+], even when the forcing frequency
exactly matches the predicted frequency. Excitation of [2,2,+]
would require two adjacent walls to move in opposition of phase,
which is not possible with a rigid container.

These observations confirm that the libration forcing is able to
excite only a subset of the possible inertial modes. Excitable
modes $[n,m,s]$ have even $n$ and $s=+$, and must in addition be
symmetric under rotation of $\pi/2$ about $z$, which is satisfied
for [2,1,+] but not for [2,2,+].

\section{Resonance curve}
\label{sec:resonance}

\begin{figure}
    \centerline{\includegraphics[width=9cm]{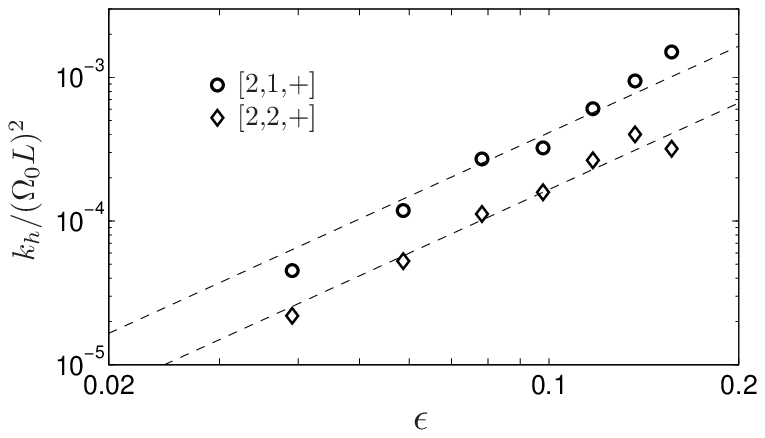}}
    \caption{Normalized kinetic energy $k_h/(\Omega_0L)^2$ of the
    wave component of the flow in the horizontal plane $z=4L/3$ as a function of
    the excitation amplitude $\epsilon$, for libration frequencies
    corresponding to
    the [2,1,+] ($\circ$) and [2,2,+] ($\diamond$) modes. The dashed
    lines show the best fits in $\epsilon^2$.}
    \label{fig:epsilon}
\end{figure}

We finally describe the energy of the oscillating flow as the
libration amplitude $\epsilon$ and the forcing frequency $\sigma_0
/ 2\Omega_0$ are varied. First, we have varied the libration
amplitude $\epsilon$ for the two frequencies $\sigma_0/2\Omega_0 =
0.648$ and $0.674$ corresponding to the (excitable) mode [2,1,+]
and the (non excitable) mode [2,2,+]. The energy of the wave
component ${\bf u'}(x,y,t)$ of the flow [see
Eq.~(\ref{eq:umulib})] is plotted in Fig.~\ref{fig:epsilon} as a
function of $\epsilon$ for these two frequencies. This energy is
computed over the horizontal components of the velocity in the
horizontal plane $z=4L/3$,
\begin{eqnarray}
k_h &=& \overline{\langle  u_x^{'2} +u_y^{'2} \rangle}_{x,y},
\end{eqnarray}
where the overbar stands for time average.

For both frequencies, the scaling of the energy is compatible with $\epsilon^2$, confirming
that the wave component of the flow essentially responds linearly
to the excitation amplitude $\epsilon$ in the range  $0.02-0.16$.
The energy for the frequency corresponding to [2,1,+] is a factor
2.5 above that for [2,2,+]. Since we know that the mode [2,2,+] is
not excited, the curve for [2,2,+] essentially corresponds to the
interfering propagating edge beams. The additional amount of
energy in [2,1,+] therefore corresponds to the resonant mode
[2,1,+]. These curves suggest that, in the horizontal plane, the energy
of the mode [2,1,+] and that of the edge waves are of the same order.

\begin{figure}
    \centerline{\includegraphics[width=11cm]{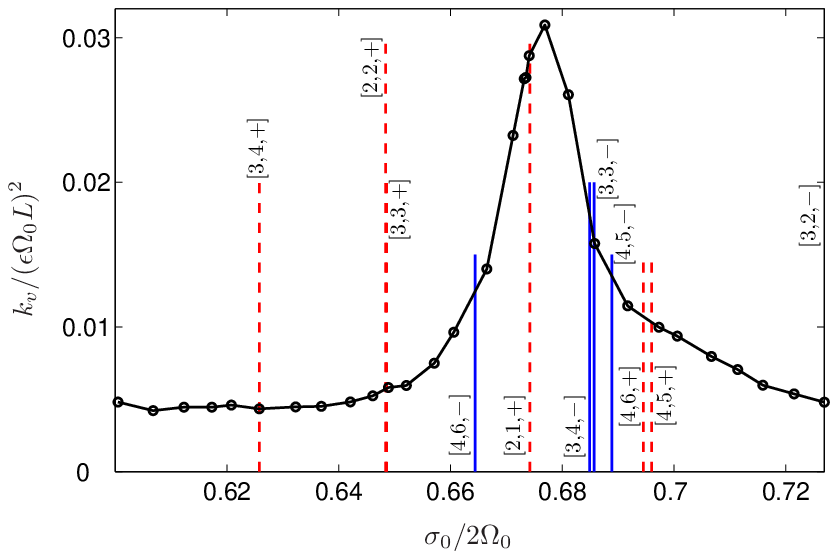}}
    \caption{(Color online) Normalized kinetic energy
    $k_v/(\epsilon\Omega_0 L)^2$ in the vertical plane $y=0$ as a
    function of the excitation frequency $\sigma_0/2\Omega_0$ for an
    excitation amplitude $\epsilon=0.02$. The vertical lines indicate
    the eigen frequencies of the inviscid modes of low order (only
    modes with $n \leq 4$ and $m \leq 6$ are shown). Solid lines
    correspond to $s=-$ modes and dashed lines to $s=+$ modes. The
    size of the vertical segments is taken proportional to $1/n$.}
    \label{fig:resonance}
\end{figure}

In order to characterize in more detail the resonance of the mode
[2,1,+], we have also computed the energy in the vertical plane
$y=0$,
\begin{equation}
k_v=\overline{\langle u_x^{2} + u_z^{2} \rangle}_{x,z}.
\end{equation}
The libration flow being normal to the measurement plane, it does not
contribute to $k_v$, so this definition essentially describes the
energy of the wave component of the flow. Focusing
on a single value of the excitation amplitude, $\epsilon = 0.02$,
we have performed a systematic scan over the excitation frequency
$\sigma_0/2\Omega_0$ in the range $[0.60; 0.73]$. Since the energy
basically scales as $\epsilon^2$, the energy $k_v$ in
Fig.~\ref{fig:resonance} is normalized now by $(\epsilon \Omega_0
L)^2$, which represents the order of magnitude
of the energy of the basic libration flow.

The resonance curve shows a well-defined peak of energy $k_v
\simeq 3 \times 10^{-2}\,(\epsilon \Omega_0 L)^2$, centered on
$\sigma_0/2\Omega_0=0.674\pm 0.005$, and of width at half maximum
$\Delta \sigma/2\Omega_0 \simeq 0.02$. The peak frequency
is very close to  the numerical prediction for the inviscid mode
[2,1,+], $\sigma_0/2\Omega_0=0.6742$, confirming that this mode is
the only one which is significantly excited in the frequency
range investigated here. We can note a slight shift of about
$\sigma_0/2\Omega_0 \simeq +0.002$ between the predicted and
the observed peak, which probably originates from a viscous detuning effect.
Although this detuning has not been computed for the
inertial modes of a parallelepiped, it is essentially similar to
the one found for the inertial modes of a sphere.\cite{Greenspan1968}
The frequency shift for the sphere has been actually determined analytically, and is
of order of $\sigma_0/2\Omega_0 \simeq
(0.02-0.05)\sqrt{E}$ for low order modes. This would give
$\sigma_0/2\Omega_0 \simeq 0.002-0.004$ for $E=5.3 \times 10^{-5}$,
which is consistent with the shift observed here for the cube.

The resonance peak is surrounded by a nearly constant plateau at
$k_v \simeq (5\pm 1) \times 10^{-3} (\epsilon \Omega_0 L)^2$,
which accounts for the wave beams emitted from the edges of the
container. This plateau confirms that the edge beams are always
present in the flow, and are hidden only at the particular
frequencies at which a mode is excited.
It is worth noting that the width of the resonance peak is
significantly larger than the width of the peak at $\sigma_0$ in
the temporal energy spectra of the flow (see
Fig.~\ref{fig:spectrum}). This suggests that the quality of the
resonance of the mode [2,1,+]  is not fixed by the precision of
the libration forcing, but is rather governed by the viscosity.
This is confirmed by the fact that the resonance width $\Delta
\sigma/2\Omega_0 \sim 0.02$ is of the order of $(\Omega_0
\tau_E)^{-1} \simeq 0.01$, where $\tau_E=L/\sqrt{\nu\Omega_0}$ is the
Ekman timescale.

As expected, inertial modes which do not have the symmetries of
the libration forcing (e.g. [3,4,+] or [3,3,+], with odd $n$), are
not found in the resonance curve. More interestingly, the even and
symmetric (with respect to rotation of $\pi$) mode [2,2,+] is not
found either, confirming that the antisymmetry (with respect to
rotation of $\pi/2$) of this particular mode is not compatible
with the symmetry of the libration. However, it appears that at
this frequency the system still lies in the far tail of the
[2,1,+] resonance, which probably explains the residual flow
structure associated to the mode [2,1,+] found in
Fig.~\ref{fig:oscflowstruct}(e). We can conclude that, because of
viscous effects and symmetry properties, the libration forcing
selects only one single resonant inertial mode in the range [0.60;
0.73]. The slight asymmetry in the tails of the resonance curve
might be due to the presence of the modes [4,5,+] and [4,6,+] at
frequencies slightly larger than the [2,1,+] peak. It turns out
that [4,5,+] has the $\pi/2$ symmetry, whereas [4,6,+] has the
$\pi$ symmetry only; it is therefore conceivable that a slight
contribution of the mode [4,5,+] is responsible for the asymmetric
shape of the resonance curve.

\section{Excitation of inertial modes by the edge beams}

\begin{figure}
    \centerline{\includegraphics[width=11cm]{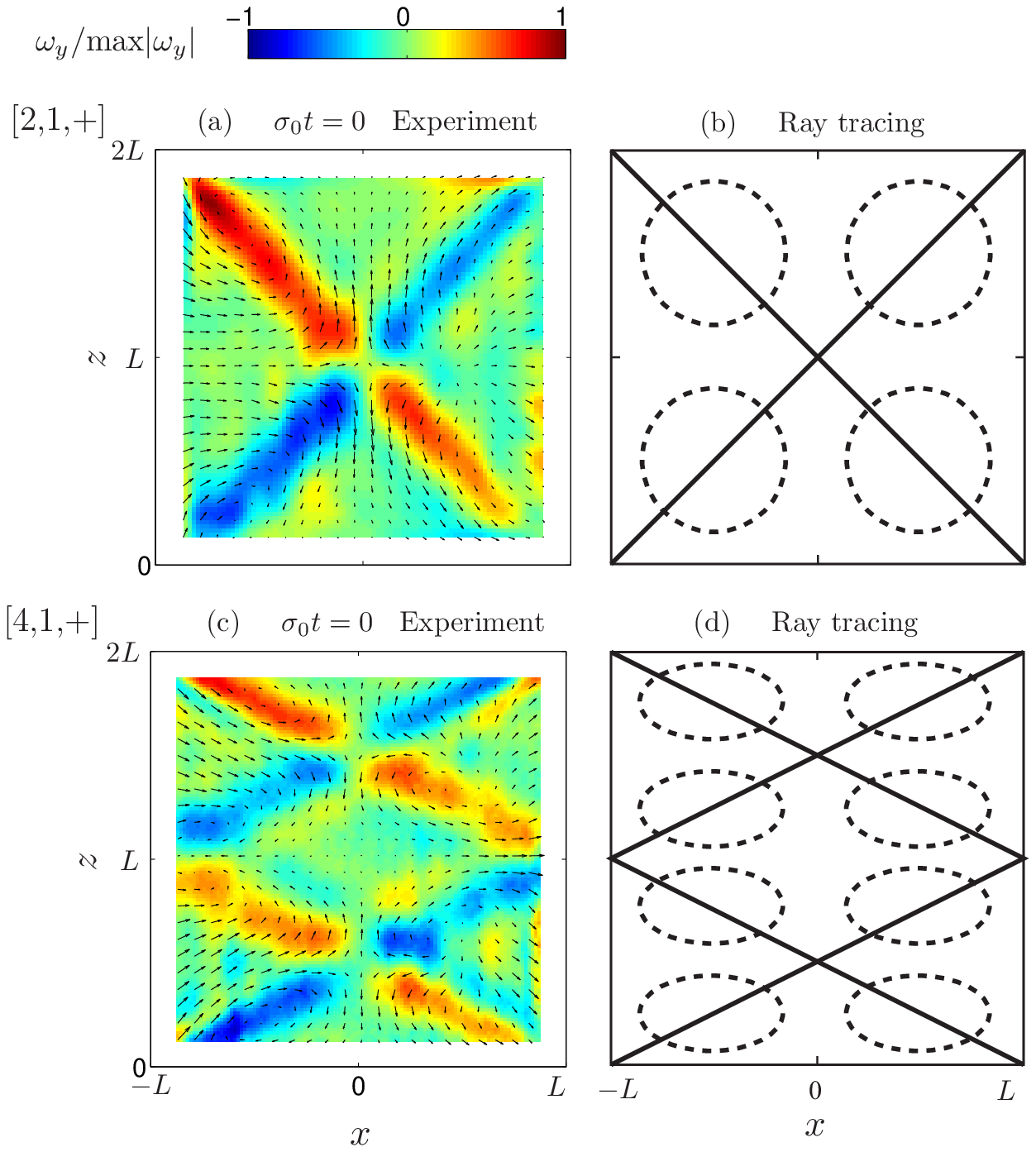}}
    \caption{(a,c) Flow in the vertical plane $y=0$ corresponding to the modes $[2,1,+]$ and $[4,1,+]$; (b,d) corresponding ray-tracing patterns (1,1) and (2,1). Figures are taken at the libration phase $\sigma_0 t = 0$, for which the mode has zero amplitude, so only the wave beams are visible.}
    \label{fig:raytrace}
\end{figure}

The present results raise the question of the relation between the
inertial modes and the edge beams originating from the convergence
of Ekman fluxes near the edges of the container.
Interestingly, for forcing frequencies close to the frequency
of low-order symmetric inertial modes $[n,1,+]$ with even $n$,
the edge beams form a periodic
ray pattern in the vertical plane $(x,z)$. This is illustrated in
Fig.~\ref{fig:raytrace}(a) and (c) for libration frequencies corresponding
to the modes [2,1,+] and [4,1,+], showing one and two
X-shaped patterns respectively. This figure suggests that periodicity of
the rays leads to the formation of a standing wave, since energy
can propagate along these rays either way. It motivates the use of a two-dimensional
simplification, in which the planar wave beams emitted from the edges
of the cube propagate in the
vertical plane along rays, at an angle given by the dispersion relation (\ref{eq:rd}). This 2D approximation does not apply near the corners,
where the flow  must be three-dimensional. But away from the walls,
the edge beams essentially protrude in the bulk of the fluid,
as if the container were infinitely long in the along-edge direction.

We consider in the following the 2D ray orbits in a vertical plane
originating from the edges of the cube.
These orbits can be characterized by  integers, $(i,j)$,
denoting the number of reflections from vertical ($i$) and
horizontal ($j)$ boundaries. Here we count edge reflections  as
reflections on \emph{both} the vertical as well as the horizontal
wall. The orbits in Fig.~\ref{fig:raytrace}(b) and (d) therefore
classify as $(1,1)$ and $(2,1)$ periodic orbits respectively. The
angle which the beam  makes with the horizontal is $
\theta_{ij} = \tan^{-1}(j/i)$, yielding a normalized frequency
$$
\frac{\sigma_{ij}}{2 \Omega_0} = \cos \left( \tan^{-1} \left( \frac{j}{i} \right) \right)=\frac{i}{\sqrt{i^2+j^2}},
$$
similar to the expression for the eigenfrequencies of the
transverse modes of an infinite channel.\cite{Maas2003} The
frequencies $\sigma_{ij}$ are dense in $[0, 2\Omega_0]$. This
spectrum is degenerate: patterns $(i,j)$ and $(i',j')$ have the
same frequency if $i'/i = j'/j$ (such degeneracy is not obvious
for the inertial modes in the parallelepiped). The resulting ray
pattern may be associated with a set of cells, with hyperbolic
points at the intersection between rays. Restricting to ray
patterns compatible with symmetries of the libration, this
construction yields $2i \times 2 j$ cells [see dashed lines in
Fig.~\ref{fig:raytrace}(b) and (d)], corresponding to an $n=2i$
inertial mode structure. No similar connection can be made between
index $m$ and $j$ from this simple geometrical construction,
because $m$ is not related to the number of cells in the
horizontal direction.

If we compare the frequencies of the periodic patterns  $(i,j)$
and the frequencies of the inertial modes $[n,m,s]$, we obtain a
reasonable agreement for $n=2i$, $m=1$, and $s=+$ (see
Tab.~\ref{tab:sij}). The discrepancy is 4.7\% for $[2,1,+]$,
and decreases for larger $n$. We can also note that, for $j>1$, it
is always possible to find, among the modes having $n=2i$ and
$s=+$, a mode $m>1$ such that $\sigma_{ij} \simeq \sigma_{nms}$ to
within a few \%. For instance, for $(i,j)=(1,2)$, one has
$\sigma_{12}/2\Omega_0 = 0.447$, which is close to the frequencies
of modes $[2,4,+]$ and $[2,5,+]$ (0.465 and 0.434, respectively).
One explanation for the apparent ability of strictly 2D periodic orbits to predict eigenfrequencies of the 3D problem quite accurately may lie in the fact that while the eigenspectrum is dense, there are indications that the density of states (the amount of eigenfrequencies per frequency increment) may be nonuniform. Computations in long channels\cite{Maas2003}  show that eigenfrequencies cluster preferentially around the eigenfrequencies of strictly transverse modes of an infinitely long channel, and there are indications that this also applies to the higher vertical modes (large $n$) in a cube.

The coincidence between the frequencies of inertial modes and
periodic beam patterns of similar spatial structure suggests the
following mechanism for the excitation of modes in a librated
container. A given libration frequency induces a set of inertial
wave beams originating from the detachment of the oscillating
Ekman layers at the edges. If the libration frequency is close to
an inertial mode, the resulting edge beam pattern is nearly
periodic, and has the same symmetries as the mode --- this is at
least the case for the two modes [2,1,+] and [4,1,+] shown in
Fig.~\ref{fig:raytrace}.  This suggests that the edge beams can
supply energy to the modes directly in the bulk of the flow, with
the correct vertical periodicity. The libration flow, being
associated to $n=0$ (vertical invariance), is unable to excite
these modes other than viscously, through the edge beams, because
it lacks the vertical structure of the inertial modes. This may be
different in the presence of sloping boundaries, which enforce
vertical motions, and may excite inertial waves inviscidly. We may
conclude that inertial modes are excited with an {\it inertial}
velocity scale $\epsilon \Omega_0 L$, but this excitation requires
the presence of edge beams induced by the {\it viscous}
oscillating Ekman layers.

\begin{table}
\centerline{
\begin{tabular}{cccccc}
\hline \hline
pattern $(i,j)$ &  $\theta_{ij}$ & $\sigma_{ij}/2\Omega_0$ & mode $[n,m,s]$ & $\sigma_{nms}/2\Omega_0$ & Difference \\
\hline
(1,1) & $45^\mathrm{o}$ & 0.707 & [2,1,+] &  0.674 & 4.7\% \\
(2,1) & $26.6^\mathrm{o}$ & 0.894 & [4,1,+] &  0.875 & 2.1\% \\
(3,1) & $18.4^\mathrm{o}$ & 0.949 & [6,1,+] & 0.938 & 1.2\% \\
\hline \hline
\end{tabular}
} \caption{Comparison between the frequencies of the periodic edge
beam patterns $(i,1)$ and the frequencies of the inertial modes
with $[n,1,+] = [2i,1,+]$.\label{tab:sij}}
\end{table}

\section{Conclusion} \label{conclusionsec}

In \emph{symmetric} containers, i.e.  axisymmetric containers  or
containers whose side walls are either parallel or perpendicular
to the rotation axis, longitudinal libration produces an inviscid
type of fluid
response that conserves absolute vorticity. However, by friction
at bottom, top and side walls, this leads to a periodic mass flux
in  boundary layers adjacent to the horizontal walls. Convergence
and divergence of these Ekman fluxes near the edges spawn free
shear layers that are inclined with respect to the rotation
axis.\cite{Kerswell1995,Duguet2006,Swart2010}  Their angle is set
by the ratio of the modulation frequency to the Coriolis
frequency. Expulsion of Ekman mass fluxes  takes place where the
topography has the same inclination, at so-called \emph{critical}
slopes. In containers where   changes in the orientation of the
boundary occur suddenly (such as near the edges of the cube
studied here) these critically sloping  regions are ``buried'' in
these horizontal edges. The free shear layers thus spawn from the
edges, and are here termed ``edge beams''.

In a 2D approximation, these edge beams propagate along periodic
paths.  For paths that are short enough, upon a few reflections
energy folds back onto itself and an inertial mode may establish
itself: a vertically standing large-scale structure that may
either be propagating or standing (``sloshing'') horizontally. In
our experiment, symmetries, due to the way that the fluid is
forced into motion by means of libration,  allow the realization
of only a particular subset of all possible theoretical inertial
modes. Latitudinal libration (libration along an axis normal to the rotation axis) might provide 
a mechanism by means of which also the other, less symmetric modes
can be excited, e.g. the mode
[1,1,-], which consists in a single oscillating cell.

Previous studies in cylindrical domains suggest that the resonant
forcing (and, therefore, amplification) of a vertically standing
inertial mode opens the possibility of  elliptic instability
(vortex breakdown). This should be manifest as a sudden collapse
of the inertial wave mode. This has so far not been seen in our
experiments in a cube, but is not ruled out either. Another
peculiarity of the theoretical inviscid wave field in the cube is
the appearance of many horizontal scales of motion,  even for
``large-scale'' modes (i.e. having low vertical wave number and
high frequency).\cite{Maas2003} These scales are required by the
wave field in order to match the rotational (circular) to the
geometrical (square) symmetries. More detailed experimental
observation of an inertial mode is required to investigate this
property. Finally, it is of interest to examine to what extent
edge beams and inertial modes can play a role in containers whose
symmetry is broken.

\appendix

\section{The inviscid base flow in a librating cube}

In this section we derive the streamfunction describing the
inviscid flow generated by the libration of a parallelepiped
container of square cross section. The Euler equation in the
librated frame of reference writes
\begin{equation}
{\partial {\bf u} \over \partial t} +({\boldsymbol \omega} + 2
{\bf \Omega}) \times {\bf u}=-{\bf \nabla}\Big({p \over \rho}+
{1\over 2} |{\bf \Omega}(t) \times {\bf r}|^2 + {1\over 2} |{\bf
u}|^2\Big)-{d{\bf \Omega}\over dt} \times {\bf r},
\label{eq:euler}
\end{equation}
with ${\bf \Omega}(t)$ the total angular velocity, of vertical
component given by Eq.~(\ref{eq:libration}). This equation
contains three terms related to the (modulated) background
rotation:  (1) the Coriolis acceleration (last term on left-hand
side), (2) the centrifugal force (second term on right-hand, that
can be absorbed in a reduced pressure) and (3) the Euler force
(last term on right-hand side), due to the modulation of the
rotation rate.\cite{Tolstoy1973} Here ${\boldsymbol \omega}$ is
the relative vorticity vector, ${\boldsymbol \omega}={\bf \nabla
\times u}$. Taking the curl of the Euler equation
(\ref{eq:euler}), we obtain the vorticity equation:
$$
{\partial {\boldsymbol \omega}\over \partial t}+ {\bf u} \cdot
\nabla ({\boldsymbol \omega} + 2 {\bf \Omega}) - ({\boldsymbol
\omega} + 2 {\bf \Omega}) \cdot \nabla {\bf u} = - 2{d
{\boldsymbol \Omega} \over dt}.
$$
This equation tells that relative vorticity changes due to
advection of absolute vorticity,
$$
{\boldsymbol \omega_a}= {\boldsymbol \omega}+2{\bf \Omega},
$$
(second term on left-hand side), vortex tilting by sheared motion
(third term), or by changes in the rotation rate (right-hand
side).

Note that for forcing by libration, the driving term on the right
of the vorticity equation is  present in the vertical direction
only and is independent of the vertical coordinate. Since the
boundaries are not inclined relative to the rotation axis this
motivates looking for velocity fields that are $z$-independent as
well. In fact, the vanishing of the vertical velocity at the
horizontal bottom and top walls suggests looking for solutions of zero
vertical velocity. The
vertical component of the vorticity  $\omega_a = {\boldsymbol
\omega}_a \cdot {\bf e}_z$ (which at this point is the only
nonvanishing component) thus satisfies
$$
{\partial \omega_a \over \partial t}=0.
$$
Since the horizontal velocity field is nondivergent ($\partial_x
u_x + \partial_y u_y =0$), we can introduce a streamfunction such
that $(u_x,u_y)=(-\partial_y \psi, \partial_x \psi)$. The spatial
part $\Psi(x,y)$ of streamfunction, $\psi(x,y,t)=\Psi(x,y)
\cos(\sigma_0 t)$, therefore satisfies a Poisson equation,
\begin{equation}
\Delta \Psi = -2 \epsilon \Omega_0,
\label{eq:apsi}
\end{equation}
subject to the condition that in the librating frame the solid
boundary at $x=\pm L$ and $y=\pm L$ acts as a streamline. We
introduce the normalized coordinates $(\tilde x, \tilde y) =
(x,y)/L$. The solution of Eq.~(\ref{eq:apsi}) is the sum of a
particular solution, $\Psi_p$, and a homogeneous solution,
$\Psi_h$, such that $\Delta \Psi_h = 0$. A particular solution is
$$
\Psi_p= \epsilon \Omega_0(1- \tilde x^2),
$$
that matches the right hand forcing term and satisfies the
boundary condition at $\tilde x=\pm 1$, but  is
\emph{nonvanishing} at $\tilde y=\pm 1$. The homogeneous solution,
$\Psi_h$, also vanishes at $\tilde x=\pm 1$, but  annihilates
$\Psi_p(\tilde x)$ at $\tilde y=\pm 1$. It is given by
$$
\Psi_h= -4 \epsilon \Omega_0 \sum_{n=0}^{\infty} (-1)^n{\cos(d_n
\tilde x) \cosh(d_n \tilde y) \over d_n^3 \cosh(d_n)},
$$
where
$$
d_n\equiv (2n+1){\pi\over 2}.
$$
For convenience, we finally symmetrize the solution, by simply
computing $(\Psi(\tilde x,\tilde y) + \Psi(\tilde y,\tilde x))/2$,
yielding
\begin{equation}
\Psi=\epsilon\Omega_0 \Big(1-{1\over 2}(\tilde x^2+\tilde y^2) -2
\sum_{n=0}^{\infty} (-1)^n{\cos(d_n \tilde x) \cosh(d_n \tilde y)
+ \cos(d_n \tilde y)\cosh(d_n \tilde x)\over d_n^3
\cosh(d_n)}\Big). \label{eq:psisol}
\end{equation}

\acknowledgments

We acknowledge A. Aubertin, L. Auffray, C Borget and R. Pidoux for
experimental help, and Y. Duguet, W. Herreman and M. Rabaud for
fruitful discussions. J.~B. is supported by the ``Triangle de la
Physique''. This work is supported by the ANR through grant no.
ANR-2011-BS04-006-01 ``ONLITUR''. The rotating platform
``Gyroflow'' was funded by the ``Triangle de la Physique''.

\end{document}